\documentclass{webofc}

\usepackage[varg]{txfonts}   
\usepackage{hyperref}
\usepackage{url}
\usepackage{siunitx}
\usepackage{lineno}
\hypersetup{colorlinks=true,citecolor=blue,urlcolor=blue,linkcolor=blue}
\title{Real-time monitoring of LHCb interaction region with a fast trackless methodology}
\author{\firstname{Giulio}~\lastname{Cordova}\inst{1,3}\fnsep\thanks{\email{giulio.cordova@cern.ch}} \and
        \firstname{Elena}~\lastname{Graverini}\inst{2,3,4} \and
        \firstname{Federico}~\lastname{Lazzari}\inst{2,3} \and
        \firstname{Michael J.}~\lastname{Morello}\inst{1,3} \and
        \firstname{Daniele}~\lastname{Passaro}\inst{1,3} \and
        \firstname{Giovanni}~\lastname{Punzi}\inst{2,3}
}

\institute{Scuola Normale Superiore, Pisa, Italy
\and
           Università di Pisa, Italy
\and
           INFN Sezione di Pisa, Italy
\and
           École Polytechnique Fédérale, Lausanne, Switzerland
          }
\abstract{The increasing computing power and bandwidth of FPGAs opens new possibilities in the field of real-time processing of high-energy physics data. The LHCb experiment has implemented a cluster-finder FPGA architecture aimed at reconstructing hits in its innermost silicon-pixel detector on-the-fly during readout. In addition to accelerating the event reconstruction procedure by providing it with higher-level primitives, this system enables further opportunities. LHCb triggerless readout architecture makes these reconstructed hit positions available for every collision, amounting to a flow of $10^{11}$ hits per second, that can be used for further analysis.
In this work, we have implemented a set of programmable counters, counting the hit rate at many locations in the detector volume simultaneously. We use these data to continuously track the motion of the beams overlap region and the relative position of the detector elements, with precisions of $\mathcal{O}\left(\si{\micro\metre}\right)$ and time granularity of $\mathcal{O}\left(\si{\milli\second}\right)$. We show that this can be achieved by simple linear combination of data, that can be executed in real time with minimal computational effort. This novel approach allows a fast and precise determination of the beamline position without the need to reconstruct more complex quantities like tracks and vertices. 
We report results obtained with $pp$ collision data collected in 2024 at LHCb.
} 

\begin{document}

\maketitle
\section{Introduction}
\label{intro}
The LHCb experiment at the Large Hadron Collider (LHC) is a forward spectrometer designed to investigate CP violation and rare heavy-hadron decays~\cite{LHCb:2008vvz}.

A significant upgrade of the LHCb detector was prepared in view of the LHC Run~3, capable of supporting an event rate of \SI{30}{\mega\hertz} by means of improved granularity and renewed readout electronics~\cite{LHCb:2023hlw}. The innermost tracking detector, used to reconstruct particle collision vertices and named VErtex LOcator (VELO), has been upgraded from silicon-strip to silicon-pixel sensors~\cite{Bediaga:2013tje}.
This detector consists of 26 layers transverse to the beam axis; each layer consists of two separable modules that can move horizontally in the transverse plane.
%
These modules can be retracted up to \SI{25}{\milli\meter} from the beam axis for protection during beam injection and brought as close as \SI{5.1}{\milli\meter} to the beam for optimal data collection. Each module is made of four hybrid silicon sensors with \SI{55}{\micro\meter}$\times$\SI{55}{\micro\meter} pixels.

A new triggerless readout architecture is one of the most significant upgrades implemented by the LHCb collaboration, making all detector data read out at every collision event~\cite{LHCb:2018mlt,LHCb:2018pqv}. In each bunch crossing, sub-detector data are time-synchronized and first grouped into event fragments before being assembled for full event reconstruction. The processing pipeline involves two software-based trigger stages (HLT1 and HLT2), where high-level physics objects such as particle tracks and primary vertices are reconstructed in real-time for immediate or offline analysis.

A notable enhancement is the introduction of a two-dimensional clustering algorithm embedded on the readout boards of the VELO~\cite{Bassi_2023}. This allows particle hits to be reconstructed at the earliest stage of the data acquisition process. The availability of a rate of $\sim 10^{11}$~hits/s in the readout FPGA enables potentially interesting measurements to be performed in real-time.
In principle, the analysis of high-statistics hit samples could serve various purposes, including detector diagnostics, alignment, luminosity, and the determination of the luminous region position. In this work, we demonstrate reconstruction of the position of the interaction region, as well as the relative positioning of VELO detector elements relative to it.  These measurements are obtained based solely on the real-time flow of reconstructed clusters, without relying on particle tracks.

\section{Real-time cluster counters on the VELO}
\label{sec:counters}

\begin{figure}
\centering
\raisebox{-.5\height}{\includegraphics[width=.3\textwidth,clip]{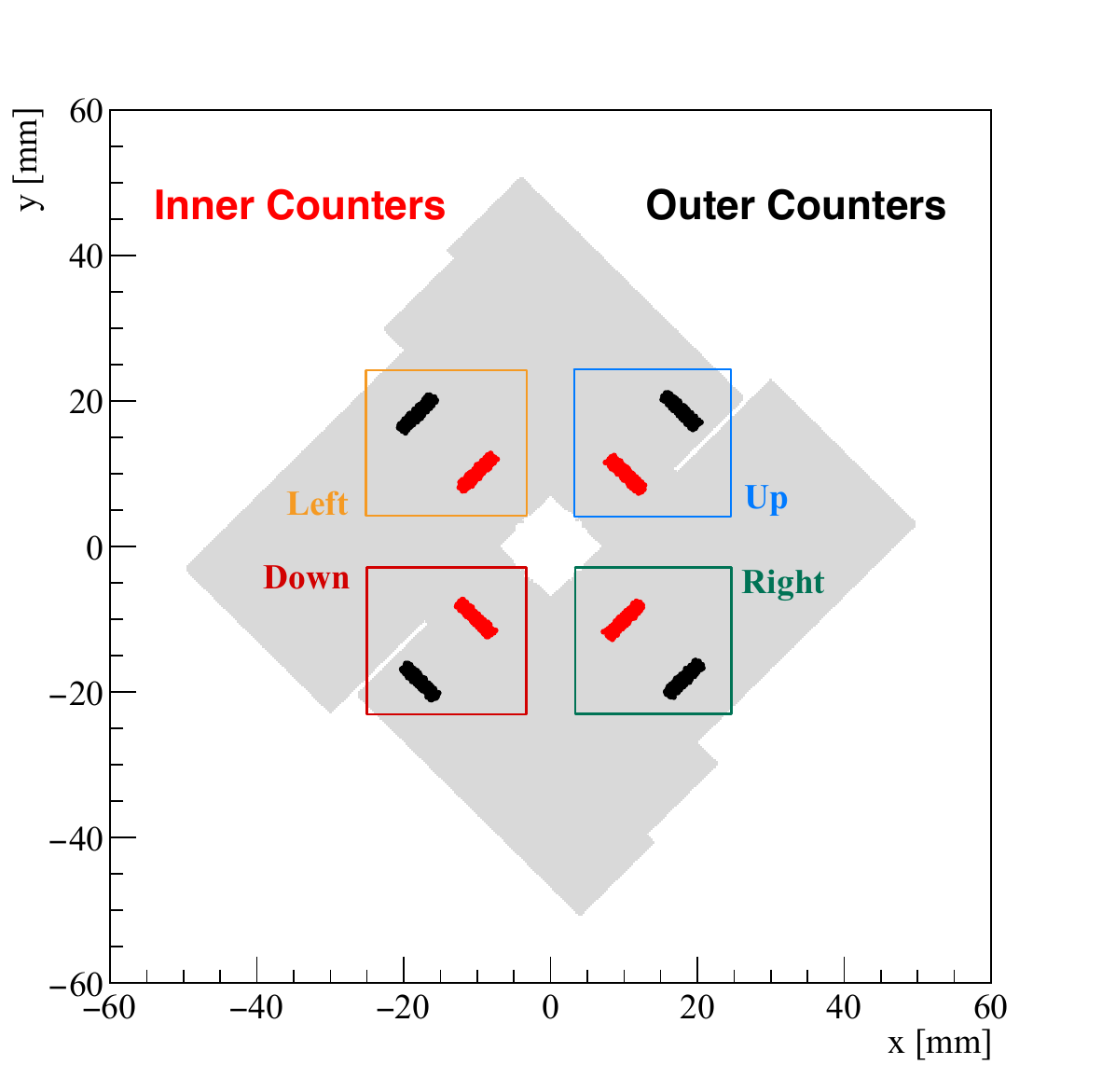}}
\raisebox{-.5\height}{\includegraphics[width=.69\textwidth,clip]{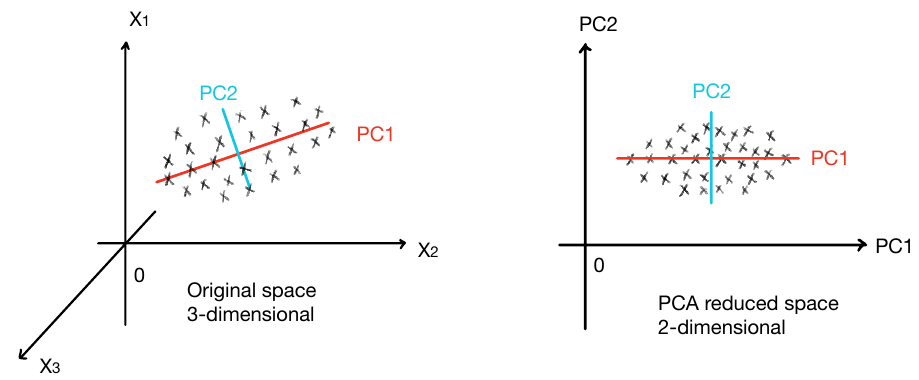}}
\caption{(Left) Position of the cluster counting regions over the VELO sensors. The colour-code highlight the position of the counters. (Right) Schematic of PCA: a 3D dataset is transformed into a 2D space using the first two principal components PC1 and PC2, ordered by variance.}
\label{fig:counters_pca}       
\end{figure}

The clustering algorithm, which is run in parallel for each VELO half-module, produces a list of $x,y$ coordinates of cluster centroids.
The latter is transmitted to the servers where partial event data are gathered together and sent off for processing by the HLT1 routines.


Two geometrical regions are defined on top of each sensor of the VELO, at different radial distances from the beam axis. Clusters recorded within these geometrical regions are accumulated to provide counters that are then used for luminosity and luminous region position analysis~\cite{LHCB-FIGURE-2024-019}. This results in a total of 208 ($2 \times 4 \times 26$) individual counters. The left-hand side of Figure~\ref{fig:counters_pca} sketches one VELO station, with the position of the counting regions highlighted in red and black for the inner and outer regions, respectively. A naming scheme is introduced to distinguish the four counting regions (Up, Down, Left and Right) within each VELO sensor. Each region is located on a different sensor of the VELO and comprises $20 \times 110$ pixels.

\section{Method}
\label{sec:method}
Information from all available cluster counters is combined to produce estimators of the position of the interaction region. Several approaches can be considered for this purpose; the method presented in this paper adopts a linear combination of the counters, where the coefficients are obtained from Monte Carlo (MC) simulated data by means of a Principal Component Analysis (PCA). 


\paragraph{Principal Component Analysis}
\label{sec:PCA}

\begin{figure}
\centering
\includegraphics[width=\textwidth,clip]{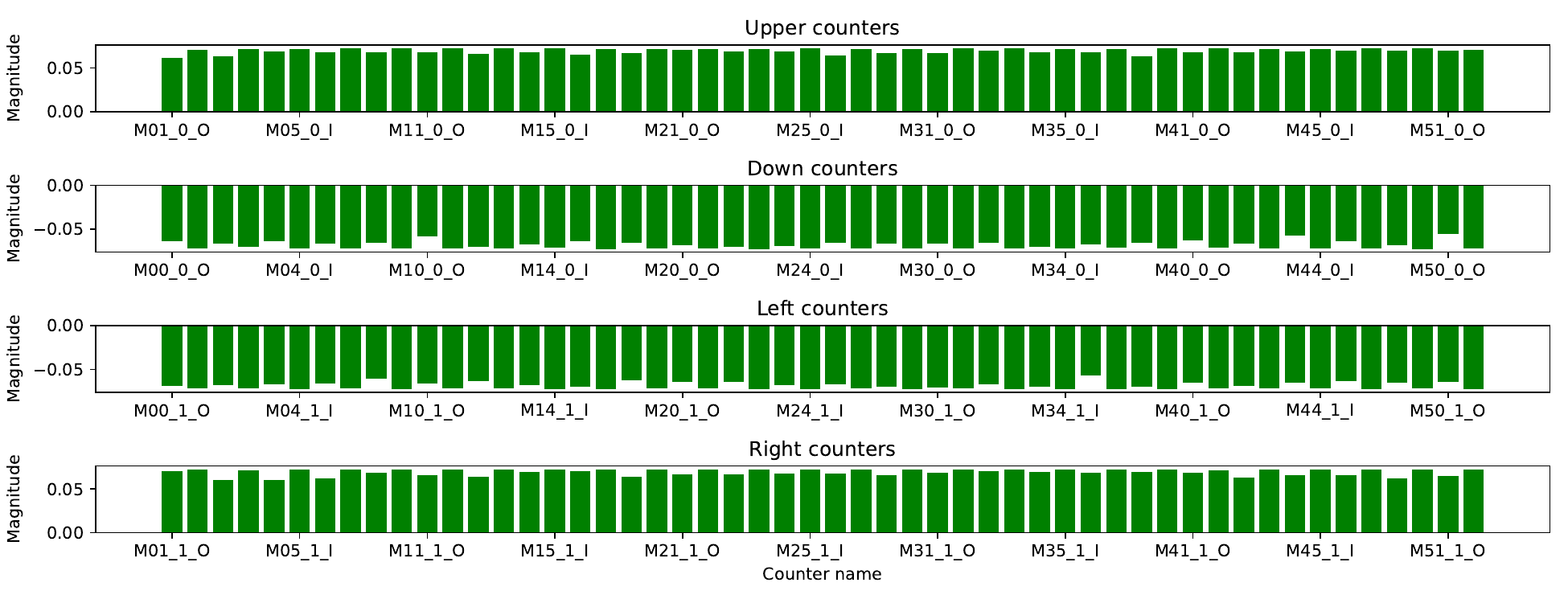}
\caption{Magnitude of each component of the eigenvectors $\mathbf{w^{x}}_{(1)}$. The elements relate to the geometrical position of the cluster counters on the VELO in this way: the first panel shows the weights $w_U$ applied to the upper counters $c_U$, the second panel shows the weights $w_L$ applied to the lower counters $c_L$, while the third and the fourth shows the weights $w_L$ and $w_R$ applied the left-hand side counters $c_L$ and right-hand side counters $c_R$, respectively. The different magnitude between the weights of outer and inner counters (even/odd columns), associated with the different acceptance, is evident. This plot also shows that the quantity $\mathbf{t^x}_{1}$ can be written in a form that highlights the weighted asymmetry up/down and right/left that is computed at each VELO layer $i$: $\mathbf{t^x}_{1}=\sum_{i} \biggl[ c^i_U \,|w^i_U| -c^i_D \cdot|w^i_D| + c^i_R \cdot|w^i_R| - c^i_L \cdot|w^i_L|\biggr]$.}
\label{fig:w1x}       
\end{figure}

Principal Component Analysis (PCA)~\cite{Pearson01111901} is a statistical technique for dimensionality reduction, feature extraction, and data compression. It transforms a dataset via an orthogonal linear transformation that maximizes variance along successive Principal Components (PCs). The right-hand panel of Figure~\ref{fig:counters_pca} provides an intuitive visualization.

Let $\mathbf{C}$ be an $n \times p$ data matrix, where the average of each column is 0. PCA finds a set of $l$ orthonormal weight vectors $\mathbf{w}_{(k)}$, mapping each row $\mathbf{c}_{(i)}$ to ordered PC scores:

\begin{equation} \mathbf{t_{k}}_{(i)} = \mathbf{c}_{(i)} \cdot \mathbf{w}_{(k)} \quad \text{for} \quad i = 1, \ldots, n \quad k = 1, \ldots, l. \label{eq:score} \end{equation}
The first PC represents the direction on which the projected dataset has the greatest variance, leading to the optimization problem:
\begin{equation}
\mathbf{w}_{(1)} = \arg \max_{\|\mathbf{w}\| = 1} \left\{ \sum_{i=1}^{n} \left( \mathbf{c}_{(i)} \cdot \mathbf{w} \right)^2 \right\} = \arg \max_{\|\mathbf{w}\| = 1} \left\{ \|\mathbf{Cw}\|^2 \right\} = \arg \max \left\{ \frac{\mathbf{w}^{\mathsf{T}} \mathbf{C}^{\mathsf{T}} \mathbf{Cw}}{\mathbf{w}^{\mathsf{T}} \mathbf{w}} \right\}.\label{argmax}
\end{equation}
This maximization corresponds to the largest eigenvalue of $\mathbf{C}^{\mathsf{T}} \mathbf{C}$, with $\mathbf{w}_{(1)}$ as its associated eigenvector~\cite{horn13}.
For subsequent components, a residual matrix $\mathbf{\hat{C}}_k$ is defined by removing the contributions of the first $k-1$ components. The problem reduces to an eigenvalue decomposition of $\mathbf{\hat{C}}_{k}^\mathsf{T}\mathbf{\hat{C}}_{k}$, ensuring orthogonality.
The PCA technique, ultimately, diagonalizes the sample covariance matrix:

\begin{equation} \mathbf{S} = \frac{1}{n-1} \mathbf{C}^{\mathsf{T}} \mathbf{C}, \end{equation}
where the eigenvectors of $\mathbf{S}$ are ordered by eigenvalue magnitude, with the largest eigenvalue corresponding to the first PC. The percentage of variance that each PC represents, corresponds to the normalized eigenvalues.

\paragraph{Parameter estimation in Monte Carlo simulation}

\begin{figure}
\centering
\includegraphics[width=.495\textwidth,clip]{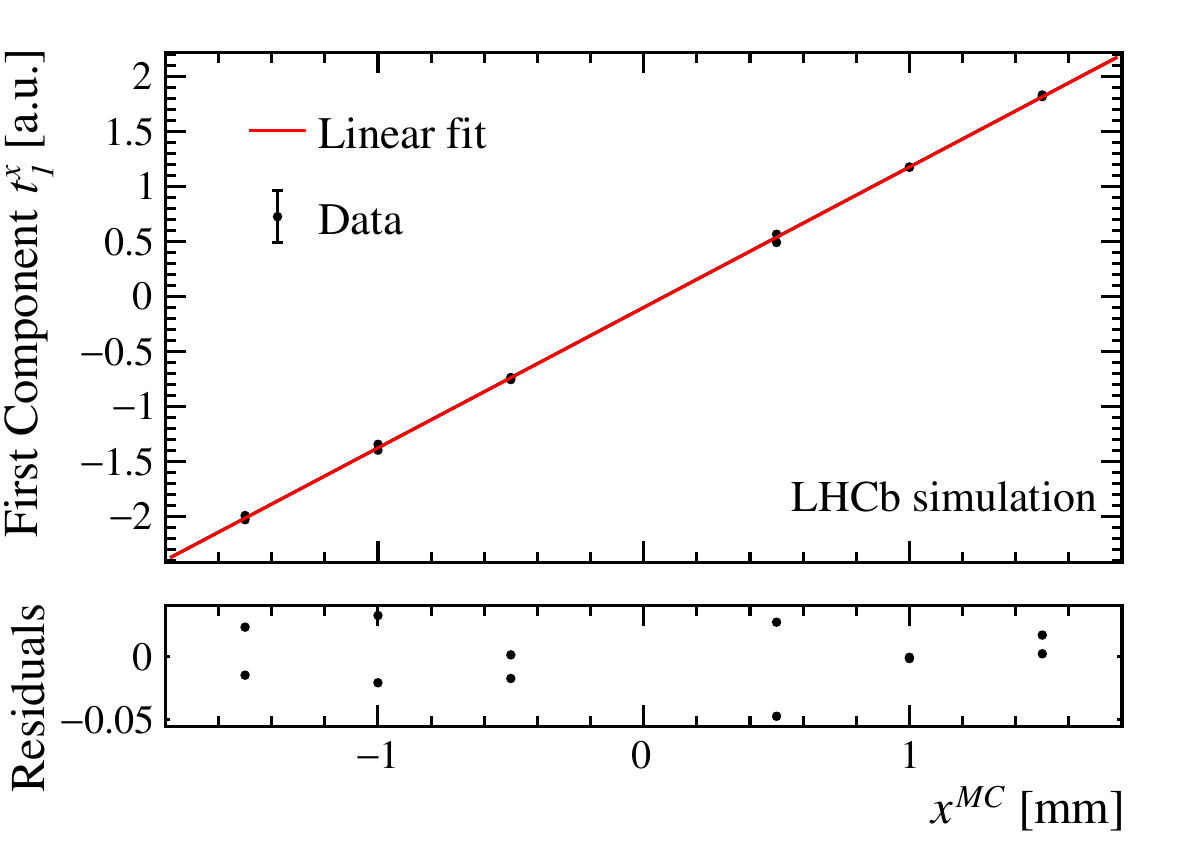}
\includegraphics[width=.495\textwidth,clip]{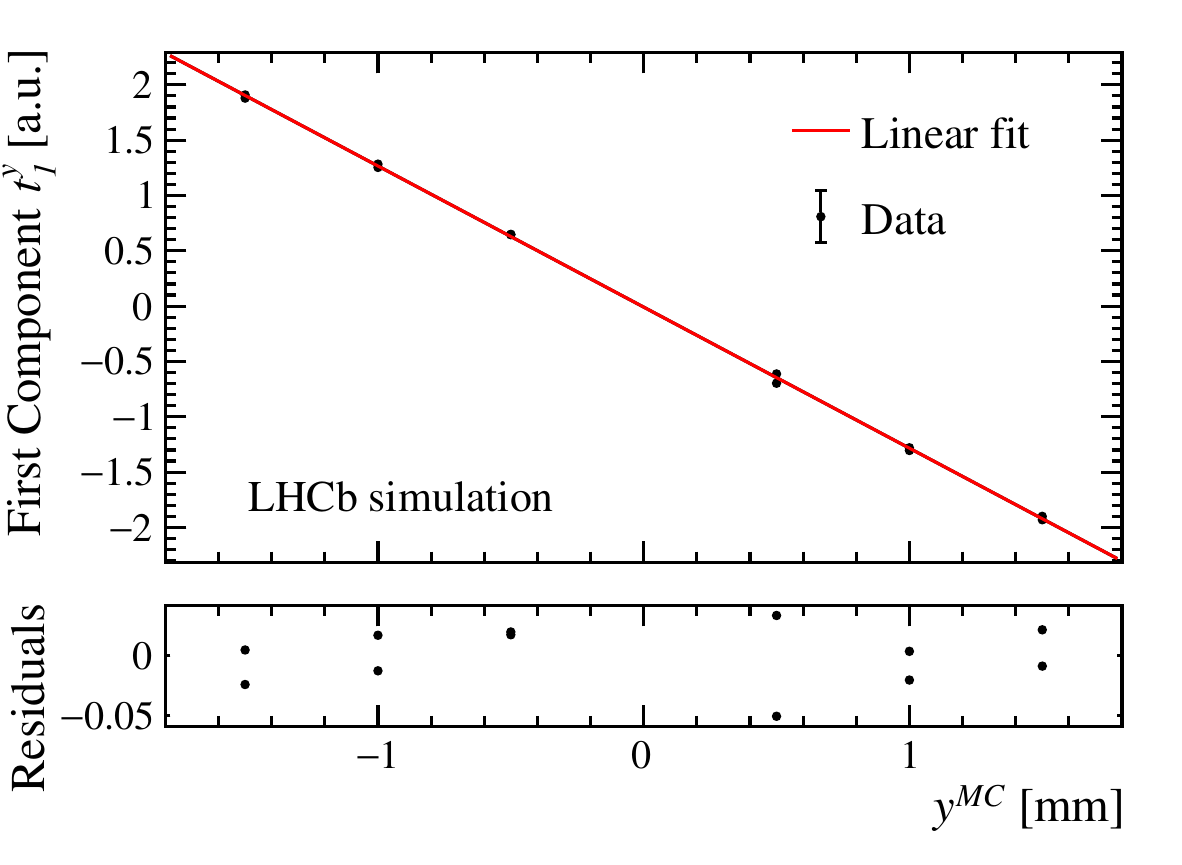}
\caption{Linearity of the first component calculated with the PCA with respect to luminous region position shifts along the $x$ (left) and $y$ directions (right).}
\label{fig:MC_x}
\end{figure}

We construct two estimators $\hat{x}$ and $\hat{y}$, to track the two transverse components of interaction region position, using a set of $l$ scores ${\left\lbrace t^j\right\rbrace}_{k=1...l}$, with $j=x,y$:
\begin{align}
    &\hat{x} = \hat{x}\left(t^x_1, \dots, t^x_l\right), 
    &\hat{y} = \hat{y}\left(t^y_1, \dots, t^y_l\right) \label{x_hat}
\end{align} 
Each of the $t_k$ scores is described by equation~\eqref{eq:score}, hence let us define $\mathbf{c}_{(i)}$ and $\mathbf{w_{(k)}}$. The $\mathbf{c}_{(i)}$ vector is a $p$-dimensional vector\footnote{where $p=208$, \textit{i.e.} the vector has one entry per counter.} containing the mean cluster count per event, normalized to the sum of each of the $p$ implemented counters. The  $\mathbf{w_{(k)}}$ is the vector of weights to calculate the $k$-th component with the PCA algorithm. Training MC datasets $\mathbf{X}_{train}$ or $\mathbf{Y}_{train}$ are used to estimate the weights. The MC datasets are generated using the LHCb simulation workflow, where the beam spot is shifted with respect to its nominal position. As described in Section~\ref{sec:PCA}, the covariance matrices of these datasets are calculated and diagonalized in order to obtain $k$  $p$-dimensional weight vectors $\mathbf{w_{(k)}}$, that represent the new basis of the space computed by the PCA. In our setup, more than 90\% of the variance of $\mathbf{X}_{train}$ and $\mathbf{Y}_{train}$ is explained by the first PC, while the others give marginal contributions.  This behaviour is expected because, in the MC training dataset, the only varying feature (i.e., the main source of variance) is the position of the luminous region. As a consequence, the first PC captures most of the variance and is expected to be directly proportional to the luminous region position, while the remaining components primarily describe statistical fluctuations in the dataset. Therefore, we will only use the first PC for each Cartesian component.
A visualisation of the components of the 208-dimensional vector $\mathbf{w^{x}}_{(1)}$ is depicted in Figure~\ref{fig:w1x}. 


In order to check the relation between $\mathbf{t^{j}}_{1}$ --the scores of the first PC-- and the luminous region position, these scores are calculated using rates $\mathbf{c_{(i)}}$ from test MC datasets. The $\mathbf{w^{j}}_{(1)}$ weights are estimated from the training datasets. Two different datasets are used in order to prevent bias. Being the test datasets completely independent from the training datasets, the validity of this method is assessed. In fact, when operating on real collision data, the scores $\mathbf{t^{j}}_{1}$ are obtained in real time from the cluster counters $\mathbf{c_{(i)}}$, and the vectors $\mathbf{w^{j}}_{(1)}$ are pre-calculated using the train MC. The relation between the first PC  and the luminous region position can be assessed in the test datasets as depicted on the left and right-hand side of Figure~\ref{fig:MC_x} for the $x$, and $y$ components, respectively.
 
The approximation of linearity is only valid for $x$ and $y$, because in this case the shifts of the luminous region are small with respect to the detector size. 
For the the $z$ direction --where the observed range is comparable to the size of the detector-- a cubic relation with $\mathbf{t^{z}}_{1}$ was observed.  Determination of the $z$ component is however of lesser importance due to wide spread of the interaction region in this coordinate , and will not be given further attention in the present work.

\paragraph{Calibration on real data}
Equation~\eqref{x_hat} can be rewritten explicitly to depend only on the first PC $\mathbf{t^j_1}$:
\begin{align}
    &\hat{x} = \hat{x}\left(t^x_1\right) = \alpha_x t^x_1 + \beta_x; 
    &\hat{y} = \hat{y}\left(t^y_1\right) = \alpha_y t^y_1 + \beta_y. \label{x_hat_true}
\end{align} 
The coefficients $\alpha_j$ and $\beta_j$ (with $j=x,y$) are obtained from a fit to calibration data, minimising the residuals of the cluster-counter based position estimators $\hat{x}$, $\hat{y}$ with respect to the position readings provided by the VELO Closing Monitoring Tasks (VCMT). These positions are obtained from distributions of PVs calculated using VELO tracks sampled every few milliseconds. 

\begin{figure}
\centering
\includegraphics[width=6cm,clip]{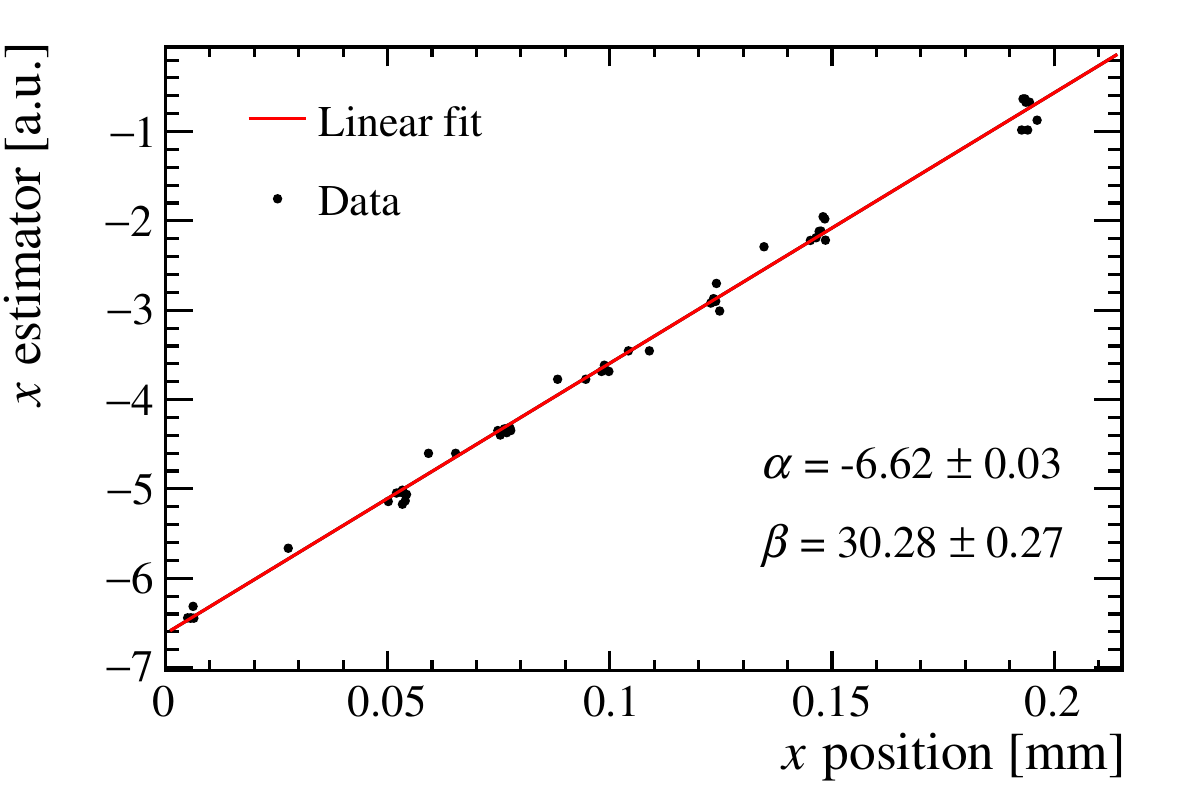}
\includegraphics[width=6cm,clip]{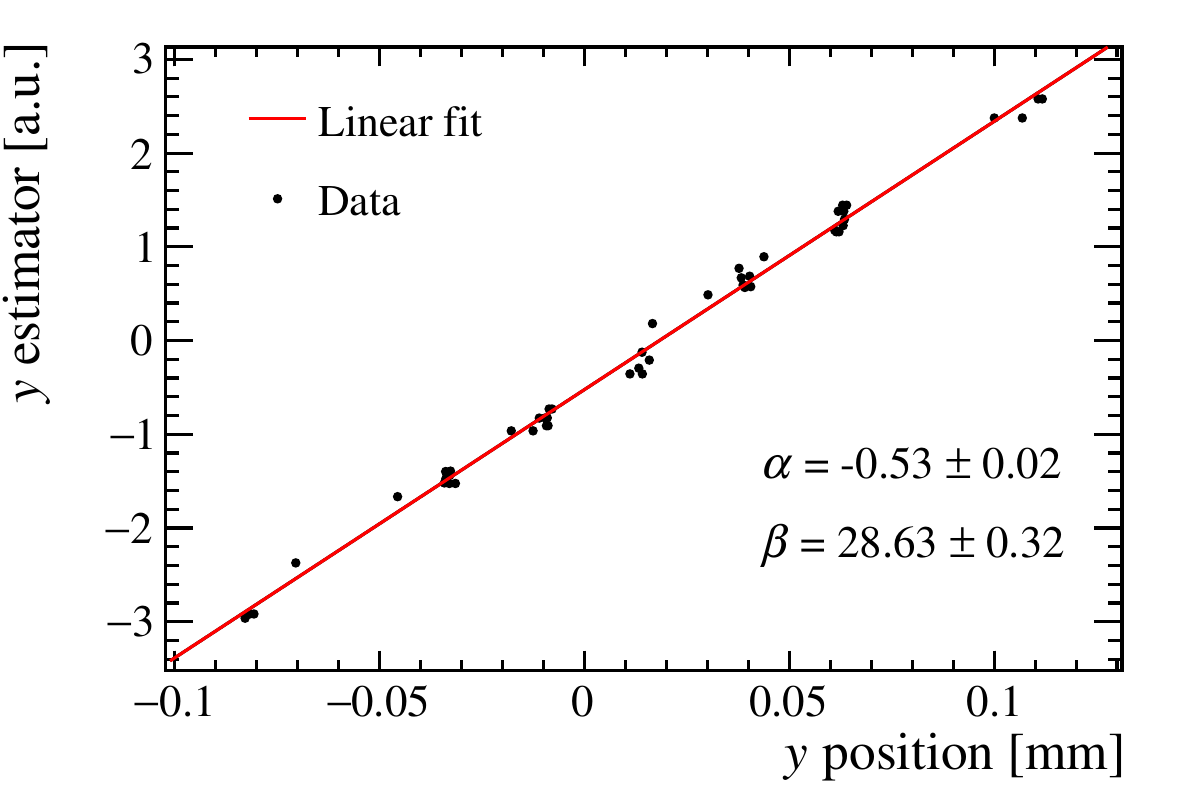}
\caption{Calibration line for the estimator of the interaction region $\hat{x}$ position (Left) and $\hat{y}$ (Right). The position shifts due to the LSC steps performed during the short vdM scan of Fill 9475 of April, 6th 2024. The $y$-axis represents the score $t^j_1$ for $j=x,y$, while the $x$-axis shows the position measured by the VCMT. A linear fit is performed to estimate $\alpha$ and $\beta$ of Equation \eqref{eq:score}. }
\label{fig:calib_on_data}       
\end{figure}

The length-scale calibration (LSC) phase of a short van der Meer (vdM)~\cite{Balagura:2020fuo} scan performed on April 6\textsuperscript{th}, 2024 is used as calibration dataset. During a LSC, the beams are shifted head-to-head in order to calibrate the size of the vdM displacement steps. For the purpose of this study, the LSC is useful because the interaction region is shifted and the estimator presented in this paper can be calibrated and tested. 
The left and right panels of Figure~\ref{fig:calib_on_data} show the calibration lines of the $x$ and $y$ estimators, respectively. 



On real collision data, we have to account for the possibility of missing counters or outliers, which could affect the computation of $t_1^j$, since this requires exactly $p$ components. Let us define $m^r(\tau)$ as the median value of the counters $c_i^r(\tau)$ at a specific time $\tau$, where the superscript $r$ denotes either inner or outer counters. If a counter value $c_i^r(\tau)$ falls outside the range $m^r(\tau)\pm 50\%$, it is marked as an outlier and replaced with the median value $m^r(\tau)$ when computing $t_1^j$. This procedure ensures the robustness of the method under experimental conditions where some counters may be missing or affected by detector or data acquisition glitches.

\section{Results on \textit{pp} collision data}
\begin{figure}
\centering
\includegraphics[width=6cm,clip]{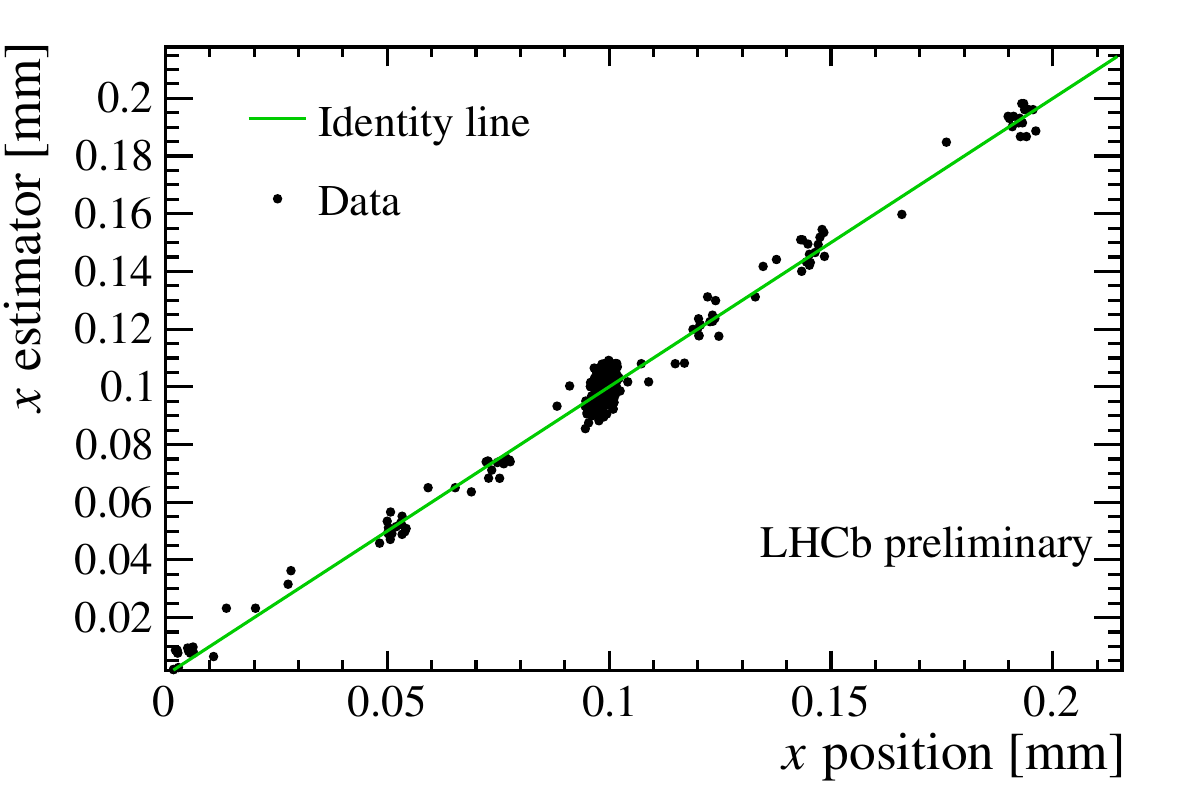}
\includegraphics[width=6cm,clip]{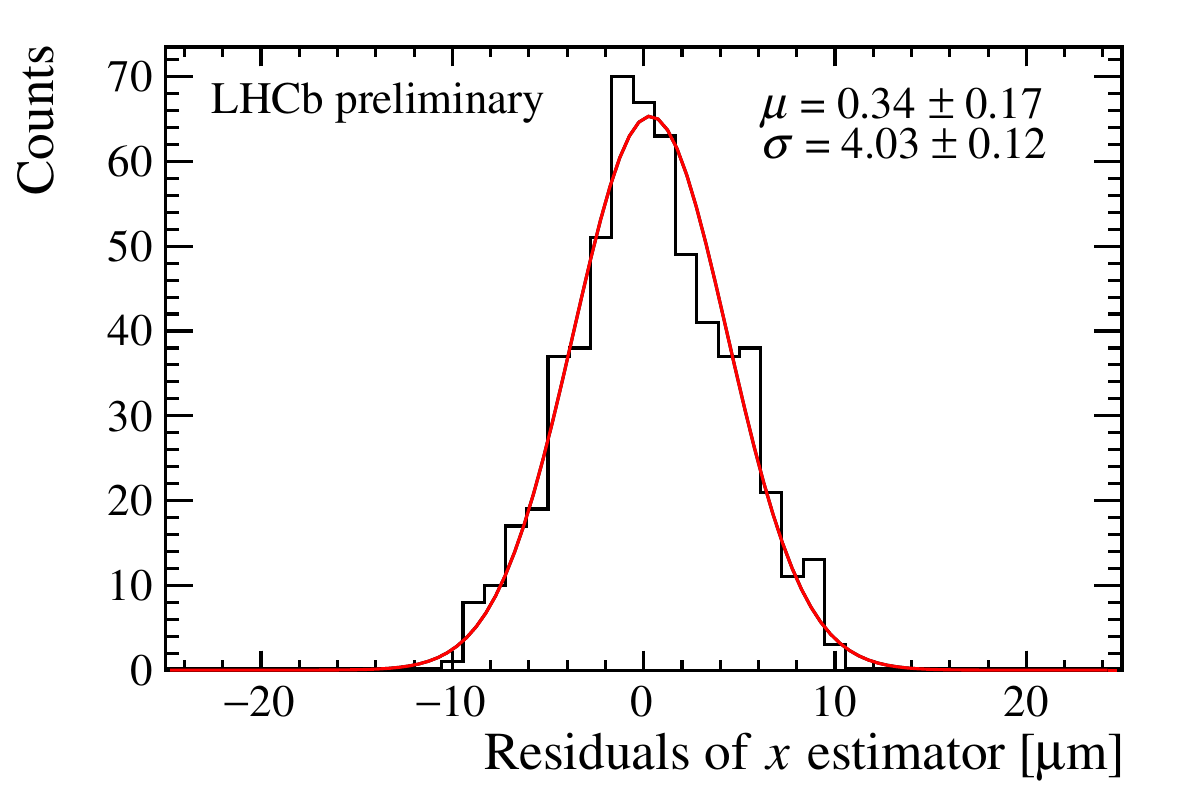}
\caption{On the left-hand panel, a comparison of the calibrated $x$ estimator [$y$-axis] with the position measured by the VCMT [$x$-axis] is shown. The plotted data correspond to a 30-minute period following the first LSC of Fill 9475, of April 6th, 2024. The green line represents the identity line. On the right-hand panel, the residuals are shown, fitted with a Gaussian distribution. The mean and standard deviation of the Gaussian indicate the bias and resolution of the estimator on a test data sample that was not used for calibration.}
\label{fig:test_on_data}       
\end{figure}
\begin{figure}
    \centering
\includegraphics[width=6cm,clip]{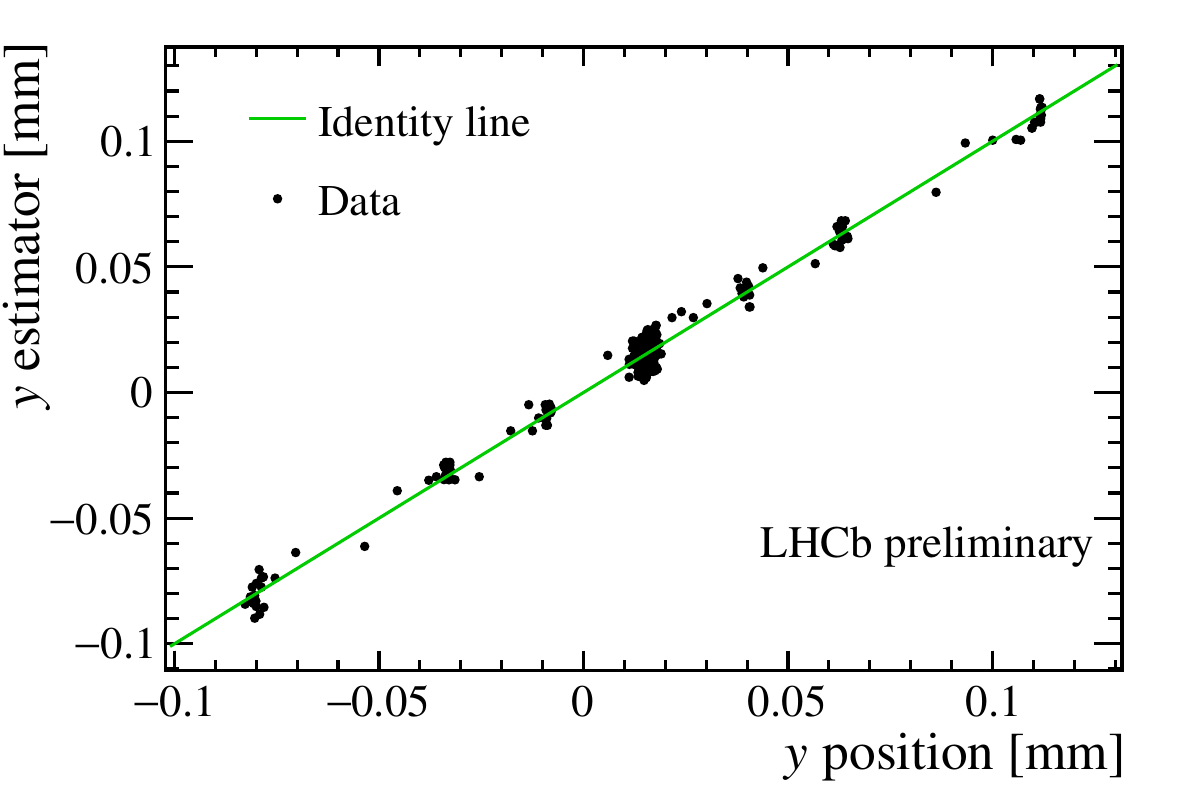}
\includegraphics[width=6cm,clip]{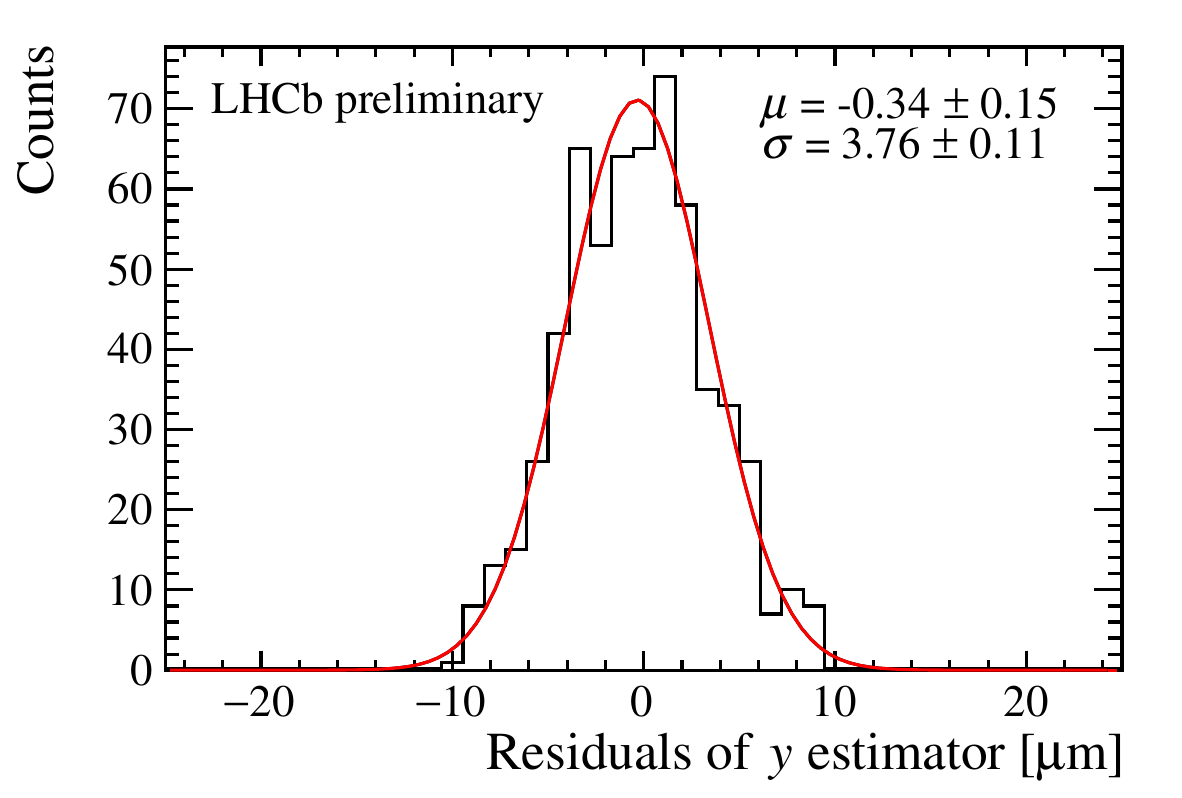}
\caption{Analogous of Figure~\ref{fig:test_on_data} for the $y$ estimator.}
\label{fig:test_on_data_y}       
\end{figure}
\begin{figure}
\centering
\sidecaption
\includegraphics[width=7cm,clip]{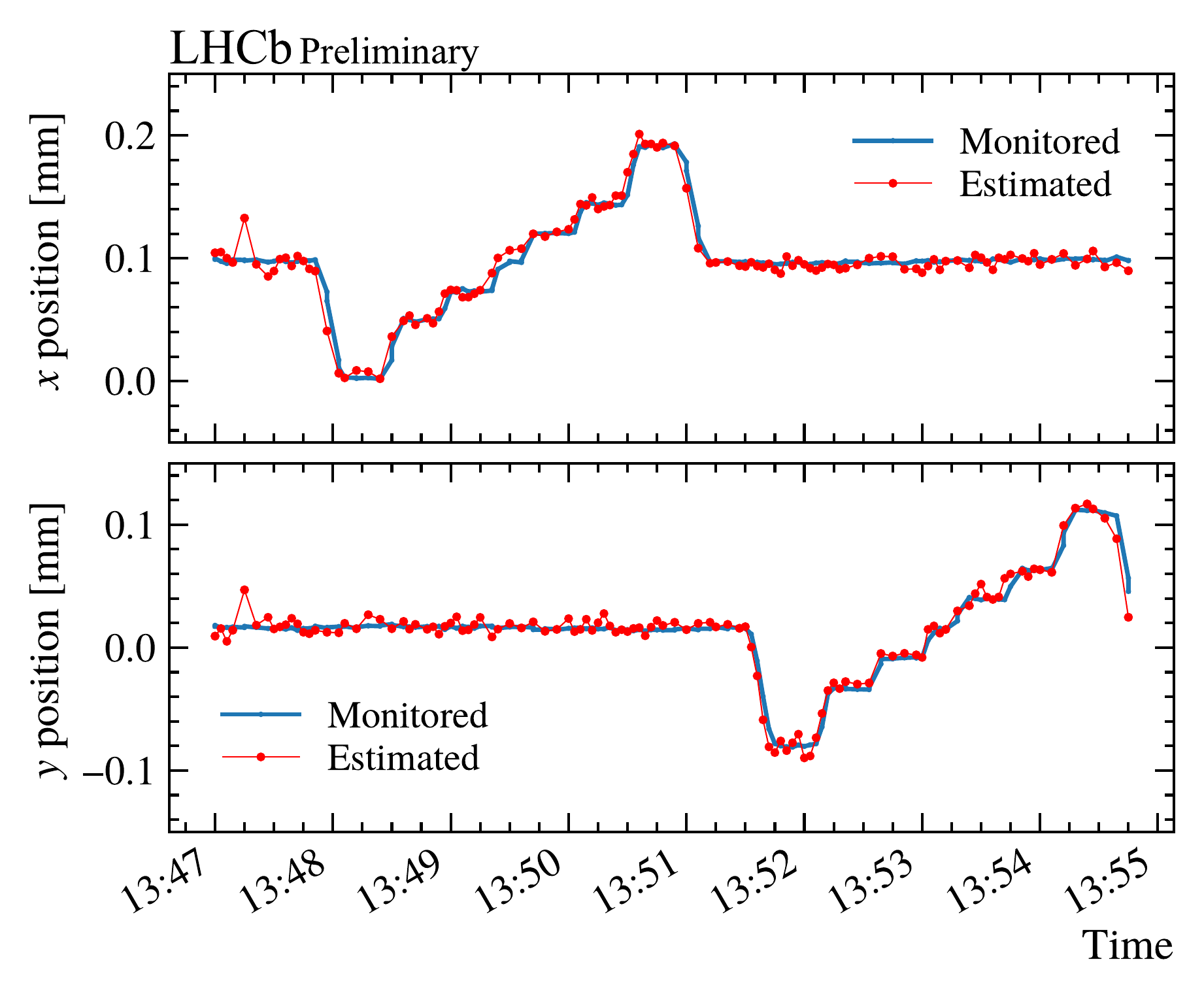}
\caption{Position of the $x$ and $y$ estimators (red) as a function of time, overlaid to the transverse beamline position provided by the VCMT (blue). The time period corresponds to a second LSC scan performed during Fill 9475 [April, 6th 2024], 30 minutes after the one used for calibration.}
\label{fig:traceplot}       
\end{figure}

The left side of Figure~\ref{fig:test_on_data} compares the estimated $x$ position versus the input from the VCMT, for a period of 30 minutes following the calibration. An identity line is overlaid to the scatter plot; the associated residuals are shown on the right-hand side of Figure~\ref{fig:test_on_data}. The same plots for the $y$ position are shown in Figure~\ref{fig:test_on_data_y}.
The estimated statistical resolution of \SI{4}{\micro\meter} is achieved accumulating counters every \SI{90}{\milli\second}. In nominal proton-proton running conditions, the same amount of counts used to compute these estimates can be achieved in just \SI{1}{\milli\second}, due to the different number of colliding bunches and different pile-up.

In Figure~\ref{fig:traceplot} we show the time evolution of the luminous region $x$ and $y$ positions as estimated by the counters presented in this work, as well as by the VCMT. 
\paragraph{Results obtained using only one half of the VELO}
\begin{figure}
\centering
\includegraphics[width=5.5cm,clip]{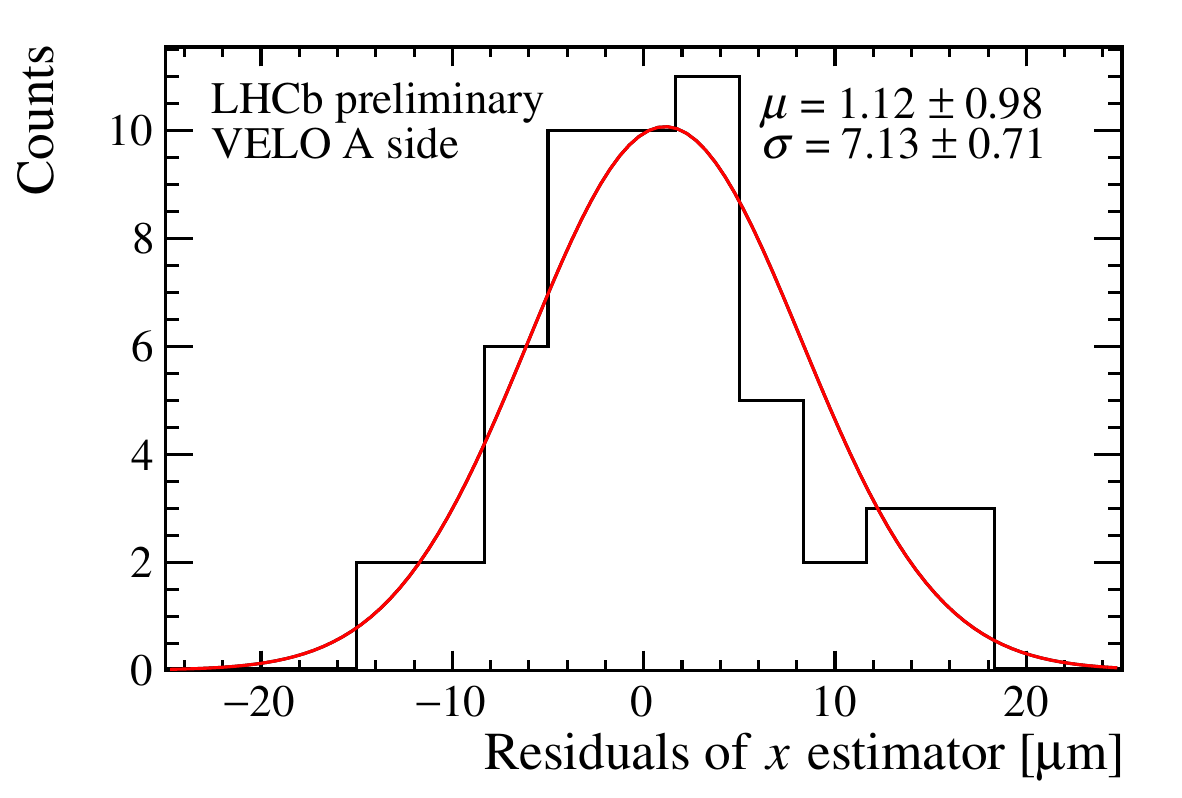}\hfil%
\includegraphics[width=5.5cm,clip]{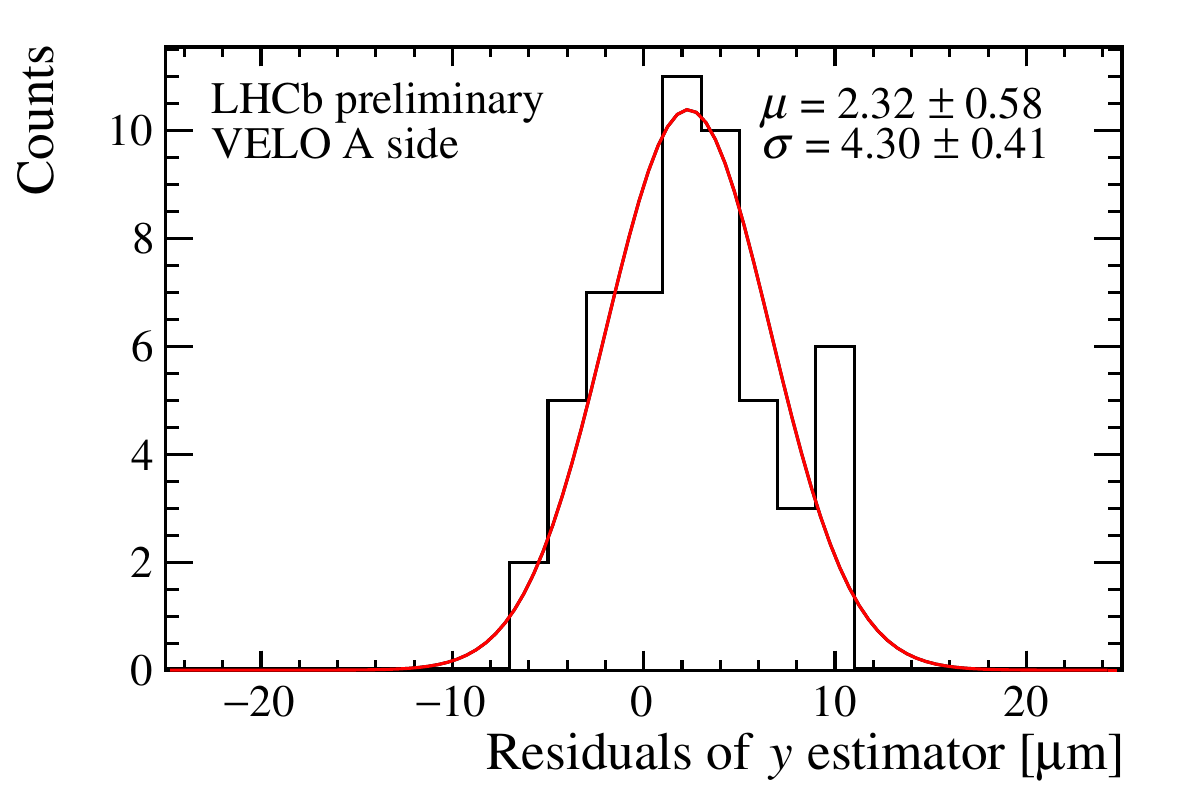}
\includegraphics[width=5.5cm,clip]{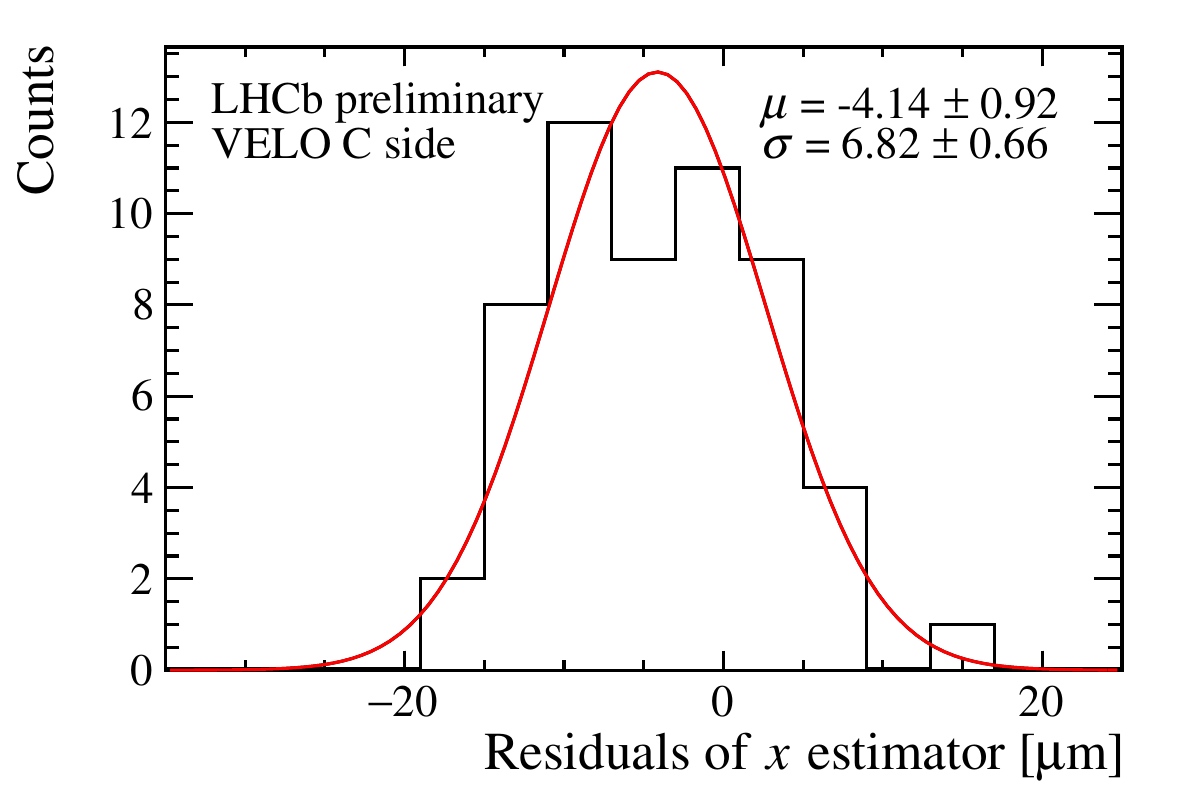}\hfil%
\includegraphics[width=5.5cm,clip]{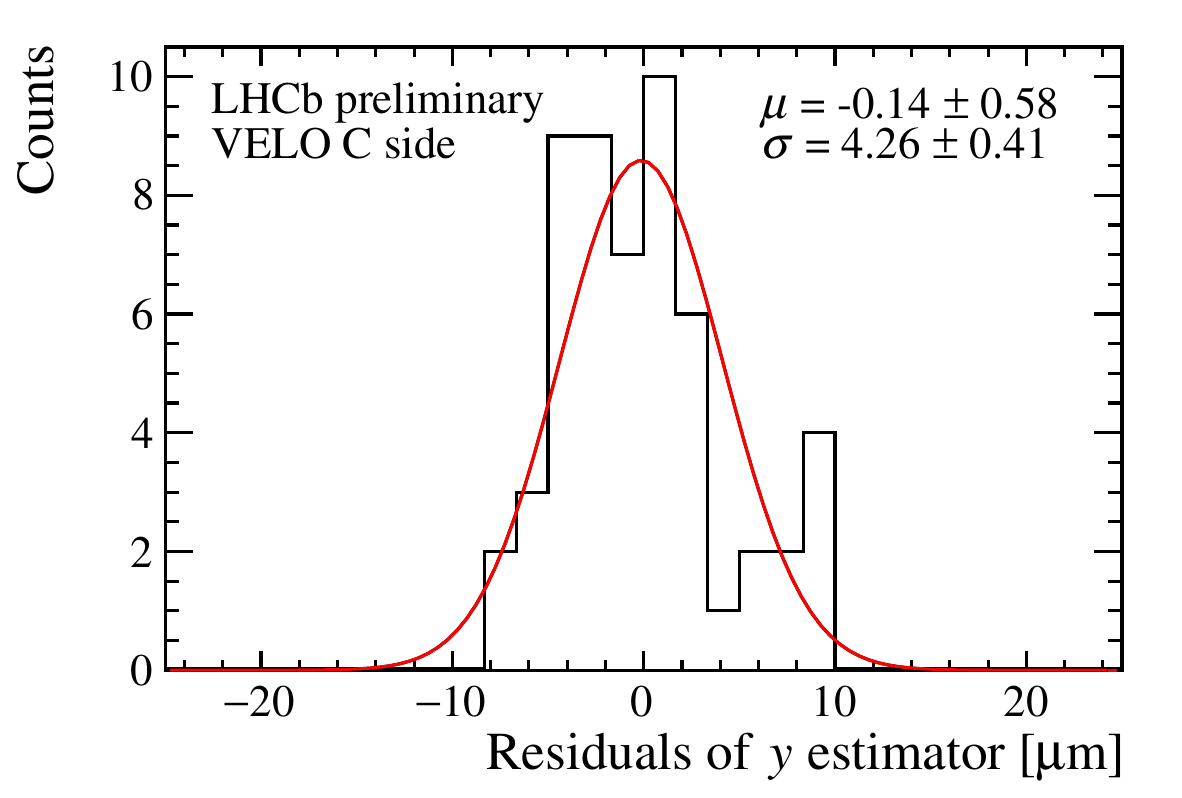}%
\caption{Residuals of estimators of the position of the VELO halves with respect to the luminous region position. (Top Left) position $x$ of VELO A. (Top Right) position $y$ of VELO A. (Bottom Left) position $x$ of VELO C. (Bottom Right) position $y$ of VELO C.}%
\label{fig:residuals_of_halves}       
\end{figure}
Alternatively, the position of the interaction region can also be estimated using only one side of the VELO. The procedure is the same as in Equation~\eqref{eq:score}, with the difference that $\mathbf{c}_{(i)}$ refers only to the counters placed on one half of the VELO, and the weights $\mathbf{w}^j_1$ are correspondingly recalculated on MC. The calibration of the VELO A-side and C-side position estimators follows the same procedure described above.

These estimators provide insight about the relative position of either VELO half with respect to the interaction region, or with respect to one another.
The VELO-half position estimates are compared with the VCMT positions, throughout a period of a few minutes following the LSC scan used for calibration.
Figure~\ref{fig:residuals_of_halves} shows the bias and resolution of these estimates for the $x$, $y$ directions and for both sides of the VELO.  Both $y$ estimators have a statistical resolution of \SI{4}{\micro\meter}, while the $x$ estimators have a statistical resolution of \SI{7}{\micro\meter}. The reason for the different resolution is still under study. 
\section{Conclusions and future prospects}
       This work focused on an application of the implementation of on-the-fly hit counters in specific VELO regions that provides real-time information embedded in the readout at the \SI{30}{\mega\hertz} collision rate. By combining these counters, we achieve a real-time trackless monitor of the luminous region position with an extrapolated statistical precision of \SI{4}{\micro\meter} using an amount of data achievable every \SI{1}{\milli\second} in nominal $pp$ running conditions. It is also possible to restrict the measurement to cluster counters from one half of the VELO. This way, we achieve a resolution of \SI{7}{\micro\meter} on the position of the luminous region; this measurement can however be interpreted also as a measurement of the VELO half position with respect to the luminous region, which could in principle prove useful for detector monitoring.
       
These results can be further improved by implementing a measurement of the $z$ component of the luminous region position, as well as a measurement of the luminous region shape and inclination in the horizontal and vertical planes.

   This study highlights the potential of real-time FPGA-based data processing, showing that it can provide reconstruction-level quantities at a very early stage in the data acquisition chain.
\bibliography{main} 
\end{document}